\documentclass[10,ptconference]{IEEEtran}
\IEEEoverridecommandlockouts
\usepackage{cite}
\usepackage{amsmath,amssymb,amsfonts}
\usepackage{graphicx}
\usepackage{algorithm} 
\usepackage{algpseudocode} 
\usepackage{quantikz}
\usepackage{textcomp}
\usepackage{xcolor}
\usepackage{caption}
\usepackage{float}

\def\BibTeX{{\rm B\kern-.05em{\sc i\kern-.025em b}\kern-.08em
    T\kern-.1667em\lower.7ex\hbox{E}\kern-.125emX}}
\begin{document}

\title{Quantum Algorithms for Optimal Power Flow
\thanks{}
}

\author{S. Fathi Hafshejani, Md Mohsin Uddin, David Neufeld, Robert Benkoczi, and Daya Gaur\\
Department of Math and Computer Science\\
University of Lethbridge, Lethbridge, AB, Canada.\\
Email: \{sajad.fathihafshejan, mdmohsin.uddin, dj.neufeld, robert.benkoczi, daya.gaur\}@uleth.ca
}

\maketitle

\begin{abstract}
% In this paper, we solve the Power Flow (PF) and Optimal Power Flow (OPF) problems by finding the search direction for convergence using quantum Harrow-Hassidim-Lloyd (HHL) and hybrid quantum-classical Variational Quantum Linear Solver (VQLS) methods. Our system demonstrates the potential advantages of quantum computing in ensuring the efficiency and accuracy of both PF and OPF solutions. To find the search direction, we applied hybrid VQLS
% algorithm in the Newton-Raphson
% method for PF. For OPF, we applied both preconditioned VQLS and preconditioned HHL to find the search direction in a variation of Interior Point Method (IPM). We compare our quantum approaches with classical approach in terms of accuracy, optimal costs and the number of iterations needed to reach the convergence point. These hybrid approaches theoretically offer scalability and faster convergence rates.

This paper explores the use of quantum computing, specifically the use of HHL and VQLS algorithms, to solve  optimal power flow problem in electrical grids. We investigate the effectiveness of these quantum algorithms in comparison to classical methods. The simulation results presented here which substantially improve the results  in \cite{neufeld2024hybrid} indicate that quantum approaches yield similar solutions and optimal costs compared to classical methods, suggesting the potential use case of quantum computing for power system optimization.
\end{abstract}


\section{Introduction}

Optimal power flow (OPF) is an idealized optimization model which minimizes the production costs of electricity subject to operational constraints. Solution to OPF can be used to adjust the power output of the generators, phase angle and magnitude of voltage to meet the load requirements of an electrical grid in a cost effective way. Some regulatory agencies state that security constrained  OPF (AC) with  unit committment should be the ultimate goal of software systems \cite{chatzivasileiadis2018lecture}. 
OPF is a nonlinear optimization problem.
%and Netwon-Raphson is a popular method for solving OPF.

There is substantial interest in solving the power flow problems using quantum computing (QC) \cite{qpf, saevarsson2022quantum, liu2022quantum, vereno2023exploiting, gao2023solving, amani2023quantum, neufeld2024hybrid, feng2023noise}. Whether any of the proposed approaches offer any quantum advantage in practice remains debatable due to read-in and read-out limitations and the presence of errors. The first detailed theoretical study that addresses whether there is any quantum advantage in solving the DC power flow problem (DCPF)  using HHL is when read-out requirements and the growth of the condition number are factored is due to \cite{pareek2024demystifying}. They find that the asymptotic complexity of solving DCPF using HHL is $O(n^{4.62} \log{n})$ where $n$ is the number of the buses, which is much larger compared to the classical method with complexity $O(n^{1.9} \log{n})$ \cite{pareek2024demystifying}. 
The study in \cite{pareek2024demystifying} raises two possible avenues for further exploring the use of QC for power flow optimization: 

{\it What if a different quantum algorithm is used in place of HHL and/or is there a way to contain the growth in the condition number?}

In this paper, we address both of the questions above. We use VQLS \cite{bravo2023variational}, a hybrid algorithm (HHL is not hybrid) in place of HHL, and we use classical preconditioning techniques \cite{saad1994ilut} to contain the growth in the condition number. Our results on solving OPF using VQLS and HHL using preconditioning indicate that there is still promise in using QC to solve OPF. A detailed theoretical study, as done for DCPF using HHL \cite{pareek2024demystifying}, is needed for OPF using VQLS. 
% We address the Power Flow (PF) and OPF problems by determining the search direction for convergence using quantum Harrow-Hassidim-Lloyd (HHL) \cite{Harrow_2009} and hybrid quantum-classical Variational Quantum Linear Solver (VQLS) \cite{bravo2023variational} methods. 
%Our approach highlights the potential benefits of QC in enhancing the efficiency and precision of OPF solutions. 
%To determine the search direction, we implemented the VQLS algorithm within the Newton-Raphson method for PF. For OPF, we employed both preconditioned VQLS ($VQLS_{p}$) and preconditioned HHL ($HHL_{p}$) to identify the search direction in a variation of the Interior Point Method (IPM) \cite{wang2007computational}. 
We compare our quantum approaches with classical approach in terms of accuracy of optimal costs and the number of iterations needed to reach the convergence point. These hybrid approaches theoretically offer scalability. The convergence rates of these hybrid quantum approaches are similar to that of classical approaches. Based on the results here, VQLS and HHL combined with preconditioning appears to be a promising direction for future study for the use of QC in solving power flow optimization problems.

\section{ Related Work}
The OPF problem has been a major focus of study in the power systems community since 1962 
\cite{carpentier1962contribution}. Various algorithmic techniques have been explored to accelerate OPF computation since then. Karmakar \cite{karmarkar1984new} introduced the classic interior point method (IPM) to solve large-scale convex optimization problems. IPM method can be used to solve OPF by solving a sequence of linear systems while maintaining feasibility within the interior of the feasible region. The Primal-Dual IPM (PDIPM) \cite{jabr2002primal} is one the variations of IPM, which simplifies the constrained optimization problem by transforming it into a Lagrangian form using the Barrier method. The Barrier method incorporates the inequality constraints into the objective function, converting them to equality constraints with a logarithmic barrier function. The first-order Karush-Kuhn-Tucker (KKT) \cite{Lange2013} conditions are used to identify a search direction given by a linear system. The Newton-Rapson method \cite{verbeke1995newton} is used to solve the resulting system of linear equations derived from the KKT conditions. PDIPM is the solver of choice for MATPOWER \cite{matpower2011zimmerman}. PDIPM, though computationally efficient, fails to ensure global convergence. The Step-Controlled PDIPM (SC-PDIPM) \cite{wang2007computational} is a variation of IPM which ensures global convergence and can be applied to solve OPF problems. SC-PDIM is the nonlinear solver of choice for this work. We explore the use of HHL with preconditioning and VQLS to determine the search direction. Integrating SC-PDIM with quantum algorithms for equation solving gives a hybrid quantum method for solving OPF, which is the object to study in this paper and addresses the limitations posed in \cite{pareek2024demystifying}. 
%The SC-PDIPM has four optimization or convergence conditions: feasibility, gradient, complementarity and cost conditions. 

Complexity of HHL is $\tilde{O}\left(\log (N) s^{2} \kappa^{2} / \epsilon\right)$ where $N\times N$ is size of the matrix, $s$ it's sparsity and $\kappa$ it's condition number.
Therefore, finding the solution of the system of linear equations $Ax=b$ using HHL scales quadricatically in the condition number of matrix $A$. The condition number \cite{higham1995condition} of a square matrix A is defined as
$\kappa(A)=\left\|A\right\|_2.\left\|A^{-1}\right\|_2$,
where $\left\|.\right\|_2$ is the spectral norm, that is, the matrix norm induced by the Euclidean norm of vectors. The condition number $\kappa$ measures how sensitive the solution is to a small change in input values.
%In numerical analysis, the condition number of matrix A describes how well or poorly the systems of linear equations $Ax=b$ can be approximated. 
If $\kappa(A)$ is small, the problem is well-conditioned; if $\kappa(A)$ is large, the problem is ill-conditioned. 
%If a matrix A is ill-conditioned, we can apply preconditioning techniques to make it well-conditioned. 
The linear systems arising from OPF suffer a growth in the condition number \cite{pglib}, so the use of HHL is ineffective \cite{pareek2024demystifying}.
Incomplete LU (ILU) decomposition \cite{saad2003iterative} is one of the popular preconditioning techniques and can be applied on the left side (left preconditioning) or the right side (right preconditioning)  of the matrix A. Incomplete LU (ILU) decomposition is a type of LU decomposition where the factors \( L \) and \( U \) are approximated by allowing only certain elements to be non-zero. ILU improves computational efficiency and maintains sparsity in the matrices. ILU is a useful preconditioner in iterative methods for solving sparse linear systems. We investigate using ILU as a preconditioning method to reduce the condition number.

%To solve PF problem, some recent works \cite{neufeld2024hybrid} utilize quantum approaches. In paper \cite{neufeld2024hybrid}, the authors applied a quantum algorithm known as Harrow, Hassidim, and Lloyd (HHL) \cite{Harrow_2009} to solve load flow problem. They analyzed their results on only two cases (case 3 and case 9Q).

\section{Optimal Power Flow (OPF)}

%\subsection{}

OPF is a nonlinear constrained optimization problem  \cite{frank2012optimal}. The objective is to optimize a certain aspect of the power system, such as minimizing generation cost, minimizing power losses, or improving voltage stability, while satisfying a set of equality and inequality constraints. These constraints typically include power balance equations, generation limits, voltage limits, and other operational limits.

\subsection{OPF Formulation}

The objective function is as follows:

\[
\min_{P,Q,V,\theta} \sum_{j} C_j\left( P_j \right)
\]

subject to the following constraints:

Power balance constraints (equality constraints):
\begin{eqnarray*}
  P_{G_i}-P_{D_i}=V_i \sum_{j \in B} V_j (G_{ij} \cos \theta_{ij} + B_{ij} \sin \theta_{ij}), \forall i \in B\\
  Q_{G_i}-Q_{D_i}=V_i \sum_{j \in B} V_j (G_{ij} \sin \theta_{ij} - B_{ij} \cos \theta_{ij}), \forall i \in B
\end{eqnarray*}
and its bound (inequality) constraints are:
\begin{eqnarray*}
    P_{G_i}^{\text{min}} &\leq& P_{G_i} \leq P_{G_i}^{\text{max}}, \quad \forall i \in G\\
    Q_{G_i}^{\text{min}} &\leq& Q_{G_i} \leq Q_{G_i}^{\text{max}}, \quad \forall i \in G\\
    V_i^{\text{min}} &\leq &V_i \leq V_i^{\text{max}}, \quad \forall i \in B\\
    \theta_i^{\text{min}} &\leq& \theta_i \leq \theta_i^{\text{max}}, \quad \forall i \in B\\
    |S_{ij}| &\leq& S_{ij}^{\text{max}}, \quad \forall (i,j) \in L
\end{eqnarray*}
where:
\begin{itemize}
    \item $C_j(P_j)$ is the generation cost of the generator $j$.
    \item $P_{G_i}$ and $Q_{G_i}$ are the active and reactive power  at bus $i$.
    \item $P_{D_i}$ and $Q_{D_i}$ are the active and reactive power demand at bus $i$.
    \item $V_i$ and $V_j$ are the voltage magnitudes at buses $i$ and $j$.
    \item $\theta_{ij}$ is the voltage angle difference between buses $i$ and $j$.
    \item $G_{ij}$ and $B_{ij}$ are the conductance and susceptance of the transmission line between buses $i$ and $j$.
    \item $B$ is the set of all buses.
    \item $G$ is the set of all generators.
    \item $L$ is the set of all transmission lines.
    \item $S_{ij}$ is the apparent power flow on the transmission line between buses $i$ and $j$.
    \item $S_{ij}^{\text{max}}$ is the maximum allowable apparent power flow on the transmission line between buses $i$ and $j$.
\end{itemize}

There are two variants of OPF that we consider in this paper.
AC Optimal Power Flow (AC-OPF) models the flow of alternating current (AC). Given the reative power in AC flows, the constraints are nonlinear, harder for to handle computationally. DC-OPF assumes direct current, voltage are assumed to be uniform and no reactive power is in the grid. These simplifications make the DC-OPF easy to solve. However, it does not model the acutal grid accurately.

\subsection{Primal Dual Interior Point Method (SC-PDIPM) for OPF}

Suppose the general form of the OPF problem is: 
\begin{align*}
    \min_{X} &f(X) \\
    &H(X) = 0 \\
    &G(X) \le 0 
\end{align*}

% $\min_{X} f(X)$

% Subject to:  $H(X) = 0$ [Equality Constraints] and $G(X) \leq 0 $ [Inequality Constraints]
%Barrier Function:

We can rewrite the problem in Lagrangian form by incorporating the inequality constraints into the objective function using a logarithmic barrier:
\begin{equation}
    L^\gamma (X, Z, \lambda, \mu) = f(X) + \lambda^T H(X) + \mu^T (G(X) +Z) - \gamma \sum_{m=1}^{ni} ln (Z_m)
\end{equation}
where
\begin{itemize}
    \item $\gamma >0 $ is barrier parameter 
    \item $Z$ is vector of positive slack variables for the inequality constraints
    \item $ni$ is the number of inequality constraints
    \item $\lambda$ and $\mu$ are Lagrangian multipliers 
\end{itemize}

%KKT Conditions:

The first order KKT conditions for the Lagrangian are:
\begin{align*}
 \nabla_x L^\gamma (X,Z,\lambda, \mu) &= 0 \\
H(X) &= 0 \\
G(X)+Z&=0\\
[\mu] Z -\gamma e &=0 \\ 
Z > 0, \mu &> 0
\end{align*}

Here,
$e$ is a unitary vector and [\ldots] diagonalizes the enclosed vector.
%Newton's Method for KKT Conditions: 
Let $L^\gamma (X,Z,\lambda, \mu)$ be $L$. By linearizing the KKT conditions around the current iterate we get:
{\footnotesize
\begin{eqnarray}\label{serach}
    \begin{bmatrix}
\nabla_{XX}^2 L & 0 & \nabla H(X) & \nabla G(X) \\
0 & [\frac{\mu}{Z}] & 0 & I \\
\nabla H(X)^T & 0 & 0 & 0 \\
\nabla G(X)^T & I & 0 & 0
\end{bmatrix}
\begin{bmatrix}
\Delta X \\
\Delta Z\\
\Delta \lambda \\
\Delta \mu 

\end{bmatrix}
=
-\begin{bmatrix}
\nabla_X L \\
\mu - \gamma [Z]^{-1} e \\
H(X) \\
G(X)+Z
\end{bmatrix}
\end{eqnarray}

}

We solve this linear system to get the search directions $\Delta X$, $\Delta Z$, $\Delta \lambda$, and $\Delta \mu$. In a hybrid quantum algorithm we determine the search direction using HHL or VQLS.

%Step Size and Optimization Variables' Update: 

The equations to determine the step size and update the other variables are as follows:

\begin{align*}
\alpha_p &= \min (\xi \min_{\Delta Z_m < 0} (\frac{-Z_m}{\Delta Z_m}), 1) \\ 
\alpha_d &= \min (\xi \min_{\Delta \mu_m < 0} (\frac{-\mu_m}{\Delta \mu_m}), 1)\\
X^{t+1} &= X^t + \alpha_p \Delta X \\
Z^{t+1} &= Z^t + \alpha_p \Delta Z \\
\lambda^{t+1} &= \lambda^t + \alpha_d \Delta \lambda \\
\mu^{t+1} &= \mu^t + \alpha_d \Delta \mu
\end{align*}
  
%Barrier Parameter Update:

We then update the barrier parameter $\gamma$:
\[
\gamma^{t+1} = \frac{\sigma (\mu^T Z)}{ni}
\]
where the constant $\sigma$ referred to as the parameter of the direction combination satisfies $0 < \sigma < 1$ and defines the trajectory to the optimal solution by combining the affine scaling direction ($\sigma =0$) and the centralization direction ($\sigma = 1$).

\subsubsection{Termination Conditions in SC-PDIPM}

SC-PDIPM has four conditions (feasibility, gradient, complementarity and cost conditions) that have to be satisfied to guarantee convergence to a solution \cite{wang2007computational}.
Out of four optimization or convergence conditions  in SC-PDIPM method, the complementarity condition is typically met the last. However, \texttt{gradcond} close to 0 is a measure of convergence as it indicates a stationary point. The  \texttt{gradcond} is given by:

\[
\texttt{gradcond} = \frac{\|\nabla_X L_t\|_{\infty}}{1 + \max \left( \|\lambda_t\|_{\infty}, \|\mu_t\|_{\infty} \right)}
\]

Where \( \nabla_X L(X,Z, \lambda, \mu) \) is the gradient of the Lagrangian with respect to \(X\).
We will use \texttt{gradcond} in the simulation studies to study the convergence of the three algorithms.

\section{Quantum Methods}
Here we introduce HHL and VQLS in brief. 
\subsection{HHL}
The quantum algorithm proposed by Harrow, Hassidim, and Lloyd known as HHL \cite{Harrow_2009} solves $Ax = b$. HHL prepares a state that is proportional to solution $\ket{x}$ and has to be read out. If read out is efficient then HHL promises a exponential speedup. For an explanation of HHL and its use in solving load flow, see \cite{neufeld2024hybrid}.
\subsection{VQLS}
 VQLS is a hybrid quantum-classical algorithm designed to solve linear systems of equations. It prepares a parameterized quantum state that satisfies the equation $A \ket{x} \sim \ket{b}$, where $A$ is the input matrix, $\ket{x}$ is the solution, and $\ket{b}$ is the result vector. VQLS minimizes the expectation value of the Hamiltonian of the system, as in the Variational Quantum Eigensolver (VQE)
 %which focuses on finding the ground state energy of quantum systems 
 \cite{peruzzo2014variational}. While VQE seeks the lowest energy state, VQLS is tailored for solving linear equations by iteratively adjusting the parameters of a quantum circuit \cite{bravo2023variational}.

To efficiently implement the VQLS algorithm, the matrix $A$ must meet certain criteria. Primarily, it should be expressible as a linear combination of unitary operators, similar to how the Hamiltonian in the VQE is represented using Pauli operators. Additionally, $A$ must be well-conditioned with a finite condition number $\kappa$, and have a bounded norm, typically \(||A|| \leq 1\).

The matrix $A$ is assumed to be expressible as a linear combination of Hermitian unitary matrices, denoted as:
\begin{equation}
A = \sum_i c_i A_i
\end{equation}
where $A_i$ are unitary matrices and $c_i$ are complex coefficients. This structure is analogous to modeling a Hamiltonian in quantum systems. 

% A typical decomposition includes the tensor products of identity and Pauli matrices, which are fundamental due to their known properties and ease of implementation in quantum gates. These matrices include:
% \begin{eqnarray*}
% I = \begin{pmatrix} 1 & 0 \\ 0 & 1 \end{pmatrix}, 
% X = \begin{pmatrix} 0 & 1 \\ 1 & 0 \end{pmatrix},  
% Y = \begin{pmatrix} 0 & -i \\ i & 0 \end{pmatrix},  
% Z = \begin{pmatrix} 1 & 0 \\ 0 & -1 \end{pmatrix}.
% \end{eqnarray*}
% These gates serve as the foundation for quantum circuit operations, making them widely used in algorithms like VQLS.

%VQLS relies on a cost function to optimize the parameterized quantum state. 
The cost function in VQLS is the residual error between the current quantum state $A|\psi(\theta)\rangle$ and the expected result $\ket{b}$. This cost function is minimized.
% Mathematically, it is expressed as $C(\theta) = \|A|\psi(\theta)\rangle - |\vec{b}\rangle\|^2$, where the inner product further simplifies this expression. 
The cost function can also be viewed as the expectation value of an effective Hamiltonian $H_G = A^\dagger(I - |\vec{b}\rangle \langle \vec{b}|)A$, which measures the distance between the current quantum state and the solution \cite{bravo2023variational}. Both local and global cost functions can be used depending on the specific problem, with global cost functions generally providing better scaling for complex systems.

% An Ansatz in quantum computing is a parameterized quantum state used in variational algorithms to approximate solutions for complex problems, such as determining the ground state energy of a quantum system. The Ansatz is typically constructed as a quantum circuit with tunable parameters, where the circuit's design plays a crucial role in the performance of the algorithm. There are various types of Ansätze, including hardware-efficient ones that are tailored to the constraints of a specific quantum device, and problem-specific Ansätze, which are designed based on the known physical properties of the system under investigation. 
In algorithms like the VQE and VQLS, the parameters of the Ansatz (parameterized quantum state) are optimized via classical computation to minimize a cost function, which brings the quantum system closer to the desired solution \cite{cerezo2021variational}. 
Algorithm 1 shows the pseduo-code for VQLS that we use.

\begin{algorithm}
\caption{Variational Quantum Linear Solver (VQLS) \cite{bravo2023variational}}
\begin{algorithmic}[1]

\State \textbf{Input:} Matrix $A \in \mathbb{R}^{2^n \times 2^n}$, Vector $b \in \mathbb{R}^{2^n}$, Number of layers $L$, Maximum iterations $M$, Convergence tolerance $\epsilon$, Step size $\alpha$
\State \textbf{Output:} Solution vector $x \in \mathbb{R}^{2^n}$
\State Compute Hermitian matrix $H_G $
\State Initialize random parameters $\theta$ for the quantum circuit

\While{The stopping condition is not met}
    \State Initialize quantum device with $n$ qubits
    \State Apply Ansatz (strongly entangling layers parameterized by $\theta$)
    \State Measure the expectation value $\langle H_G \rangle$
    \State Update $\theta$ using an optimization approach
\EndWhile

\State Find the optimal parameters $\theta^*$
\State Extract the optimal solution $x^*$ from the quantum circuit
\State \Return $x^*$
\end{algorithmic}
\end{algorithm}

\section{Methodology}

%\subsection{

%{We desctep Controlled Primal Dual Interior Point Method (SC-PDIPM) for OPF with Preconditioned VQLS/HHL}}

We modified the SC-PDIPM method described in \cite{wang2007computational} to solve OPF problem using quantum approaches as described in Algorithm 2. We applied a left preconditioning technique with VQLS/HHL (line 9) to find the search direction.% towards convergence point. 

\begin{algorithm}
\caption{Step Controlled Interior Point Method Using VQLS/HHL}
\begin{algorithmic}[1]
\State Let $0< \kappa < 1$, $0< \eta << 1$, $0< \epsilon << 1 $
\State Initialize $x_0, z_0, \lambda_0, \mu_0$ and $\gamma_0$ and set parameters $\epsilon_{\text{feas}}, \epsilon_{\text{grad}}, \epsilon_{\text{comp}}, \epsilon_{\text{cost}}$
\State $scipm \gets false$
\State $L_t = L^\gamma_t  (X_t, Z_t, \lambda_t, \mu_t)$ 
\State $\psi_t (\Delta X_t) = (\nabla_X L_t)^T \Delta X_t + 1/2 (\Delta X_t)^T ({\nabla_x}^2 L_t) \Delta X_t$

\State $ \rho_t (\Delta X_t)  = \frac{L^{\gamma_t} (X_t+\Delta X_t, Z_t, \lambda_t, \mu_t) - L_t}{\psi_t (\Delta X_t)}  $
\State Calculate $feascond$, $gradcond$, $compcond$, $costcond$
\While{$\forall cond \in (feascond,gradcond,compcond,costcond)>\epsilon$} 
	%\State Form the KKT matrix $A$ and right-hand side vector $b$
    
	%\State Apply ILU preconditioning:
%	\State \hspace{1em} $ILU = \text{spilu}(A)$
	%\State \hspace{1em} $M = \text{LinearOperator}(A.shape, ILU.solve)$
	%\State \hspace{1em} $A_{pc} = M @ A$
%	\State \hspace{1em} $b_{pc} = M @ b$
    
	%\State Solve $\Delta X_t,\Delta Z_t, \Delta \lambda_t, \Delta \mu_t$ using VQLS or HHL
  \State Solve system (\ref{serach}) using VQLS or HHL

    \If{$scipm = true $}
    	\While{$\rho_t (X_t) < 1-\eta$ or $\rho_t (X_t) > 1+\eta$} 
        \State $\Delta X_t \gets \kappa \Delta X_t$, $\Delta Z_t \gets \kappa \Delta Z_t$, $\Delta \lambda_t \gets \kappa \Delta \lambda_t$, $\Delta \mu_t \gets \kappa \Delta \mu_t$

     \EndWhile
	\EndIf
    
	\State Compute ($X_{t+1}, Z_{t+1}, \lambda_{t+1}, \mu_{t+1}, \gamma_{t+1}$)

	\If{$feascond_{t+1} \geq feascond_t $ and $gradcond_{t+1} \geq gradcond_t $}
    	\State $scipm \gets true$
	\EndIf
    
	\State $t \gets t+1$
\EndWhile
\State \Return $X, $Z$ \lambda, \mu$
\end{algorithmic}
\end{algorithm}

\section{Results and Discussion}
We used OPF test cases from \cite{matpower2011zimmerman} for simulation studies. Since the simulation of HHL and VQLS is time-consuming, we were limited to small-sized instances. Though, this is already an improvement on the simulations in \cite{neufeld2024hybrid}. The hybrid quantum algorithms converged to the solution of the same quality (as the classical approach). The preconditioning techniques lead to a controlled growth in the condition number, thereby reducing the number of iterations needed by HHL and VQLS.

Table \ref{table:DC-OPF-comparison} shows the number of iterations needed to converge for Classical, VQLS,  HHL (both with preconditioning) methods and the cost of the solution for these cases of Direct Current OPF (DC-OPF).

\begin{table}[H]
    \centering
    \begin{tabular}{|c|c|c|c|c|c|c|}
        \hline
        \textbf{Case} & \multicolumn{2}{c|}{\textbf{Classical}} & \multicolumn{2}{c|}{\textbf{$VQLS_p$}} & \multicolumn{2}{c|}{\textbf{$HHL_p$}} \\ \cline{2-7}
                      & It & Cost & It & Cost & It & Cost \\ \hline
        3        &      8    &  746.25 &    8       &   746.25 &          8 & 746.25  \\ \hline
        6ww      &     9     & 2393.31  &     10    &  2393.31 &          10 &  2393.31\\ \hline
        9        &    8     &  4131.03 &     8     &  4131.03 &          - &  -\\ \hline
    \end{tabular}

    \caption{Cost and the number of iterations for  Classical, Preconditioned VQLS ($VQLS_p$), and Preconditioned HHL ($HHL_p$) methods  for different cases in DC-OPF.}
    \label{table:DC-OPF-comparison}
\end{table}

Table \ref{table:AC-OPF-comparison} shows the number of iterations and cost for Classical and VQLS (Preconditioned) methods for only case 3 in Alternating Current OPF (AC-OPF). The simulation was not possible for larger cases because of increased size (due to additional constraints). This imposes a requirement of many additional qubits.

\begin{table}[H]
    \centering
    \begin{tabular}{|c|c|c|c|c|}
        \hline
        \textbf{Case} & \multicolumn{2}{c|}{\textbf{Classical}} & \multicolumn{2}{c|}{\textbf{$VQLS_p$}}  \\ \cline{2-5}
                      & Iterations & Cost & Iterations & Cost  \\ \hline
        3        &     10     & 758.21  &         10  &  758.21 \\ \hline
        
    \end{tabular}

    \caption{Cost and iterations for Classical and Preconditioned VQLS ($VQLS_p$) methods for AC-OPF.}
    \label{table:AC-OPF-comparison}
\end{table}

%\subsection{Results for Optimal Power Flow}

%Iteration vs Gradcond for Case 3-DC-OPF is shown in figure \ref{fig:case3_dc_opf_iteration_vs_gradcond}

\begin{center}
\begin{figure}[H]
    \centering
    \includegraphics[width=1\linewidth]{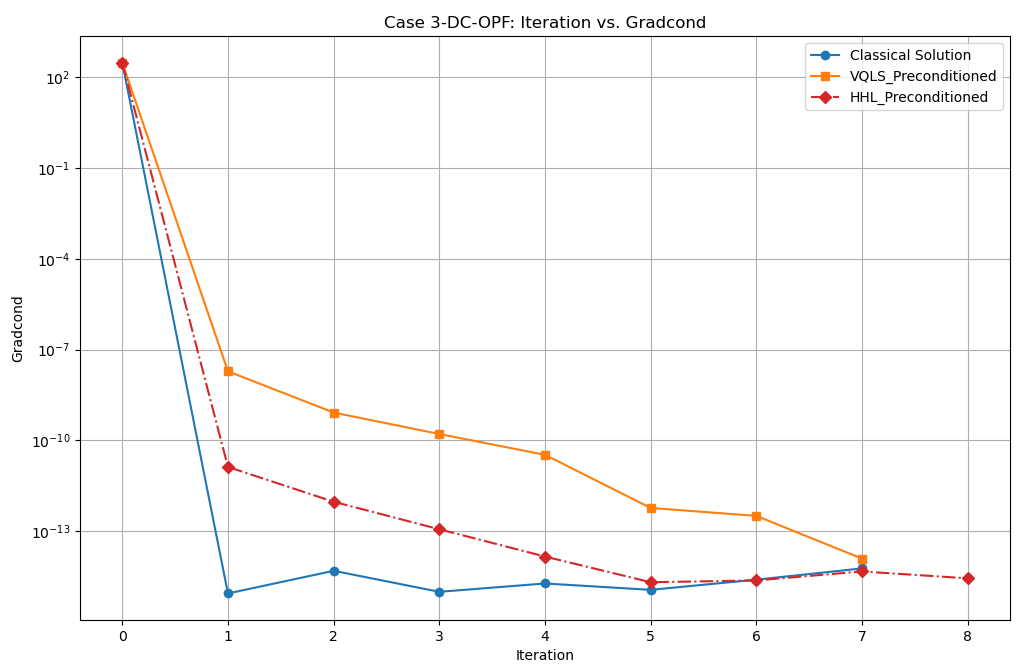}

    \caption{Case 3-DC-OPF: \texttt{gradcond} v/s iterations}
\label{fig:case3_dc_opf_iteration_vs_gradcond}
\end{figure}
\end{center}

We study the convergence of the three methods. 
Values of \texttt{gradcond} close to 0 indicate a locally optimal solution and convergence. Therefore, we plot  \texttt{gradcond} values as function of the iteration number for the cases studied here.

Figure \ref{fig:case3_dc_opf_iteration_vs_gradcond} shows the value of the \texttt{gradcond} over iterations for the classical solution, VQLS preconditioned and HHL preconditioned methods. All methods start with the same initial \texttt{gradcond} value of 300. For the classical solution, the \texttt{gradcond} values drop significantly after the first iteration. The \texttt{gradcond} values of the preconditioned VQLS and preconditioned HHL decrease steadily and rapidly over the iterations. The preconditioned VQLS and HHL demonstrate similar performance, with slight variations in their convergence patterns.

%Iteration vs Gradcond for Case 6ww-DC-OPF is shown in figure \ref{fig:case6ww_dc_opf_iteration_vs_gradcond}

\begin{figure}[H]
    \centering
    \includegraphics[width=1\linewidth]{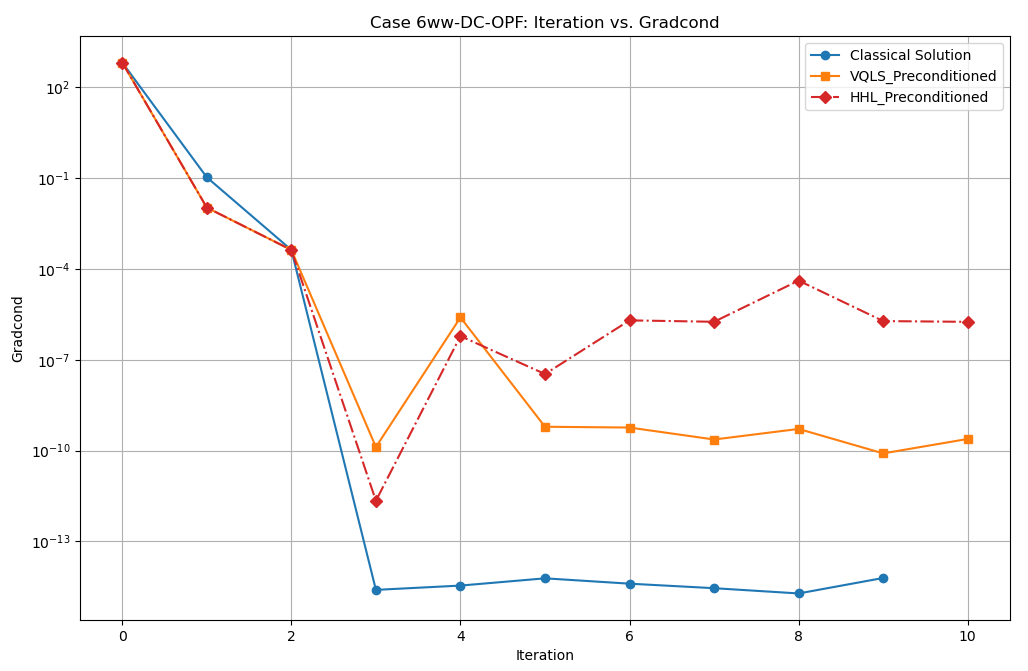}
    \caption{Case 6ww-DC-OPF: \texttt{gradcond} v/s iterations}
    \label{fig:case6ww_dc_opf_iteration_vs_gradcond}
\end{figure}

Figure \ref{fig:case6ww_dc_opf_iteration_vs_gradcond} shows the convergence behavior of the three methods. The classical solution quickly reaches very low \texttt{gradcond} values after the first iteration, and despite some fluctuations, it remains low. The VQLS preconditioned method shows a rapid decrease in \texttt{gradcond} values initially but has fluctuations in the later iterations, indicating instability compared to the classical solution. The HHL preconditioned method shows a similar rapid decrease but has more fluctuations, with a notable spike at the seventh iteration, indicating more instability than VQLS. The classical solution shows the most stable performance, while the preconditioned VQLS and HHL methods exhibit some instability, with HHL showing the most fluctuations.

%Iteration vs Gradcond for Case 9-DC-OPF is shown in figure \ref{fig:case9_dc_opf_iteration_vs_gradcond}

\begin{figure}[H]
    \centering
    \includegraphics[width=1\linewidth]{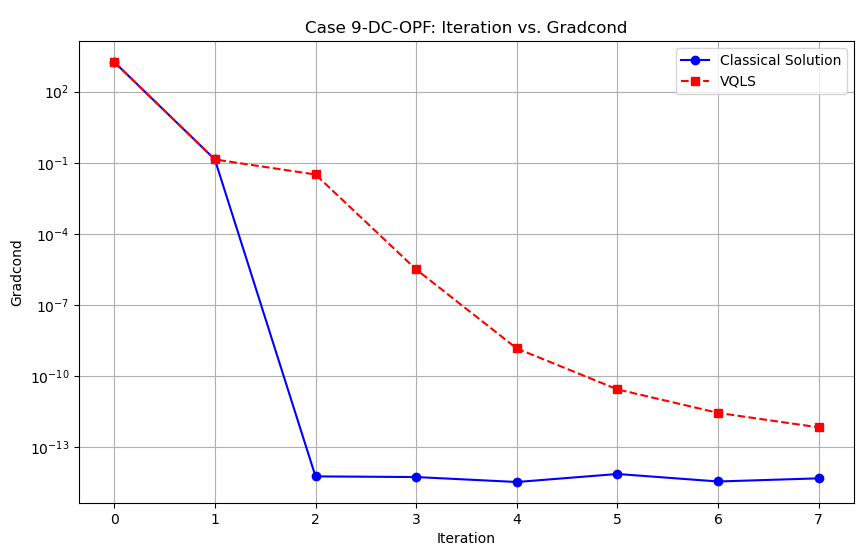}
    \caption{Case 9-DC-OPF: \texttt{gradcond} v/s iterations}
    \label{fig:case9_dc_opf_iteration_vs_gradcond}
\end{figure}

Figure \ref{fig:case9_dc_opf_iteration_vs_gradcond} compares the \texttt{gradcond}  values over iterations for the classical and VQLS preconditioned methods. The classical solution shows a rapid and significant drop in \texttt{gradcond}  after the first iteration, stabilizing at extremely low values. The VQLS preconditioned method shows a gradual decrease in gradcond, with higher values compared to the classical solution up to the third iteration before stabilizing at very low values.

%Iteration vs Gradcond for Case 3-AC-OPF is shown in figure \ref{fig:case3_ac_opf_iteration_vs_gradcond}

\begin{figure}[H]
    \centering
    \includegraphics[width=1\linewidth]{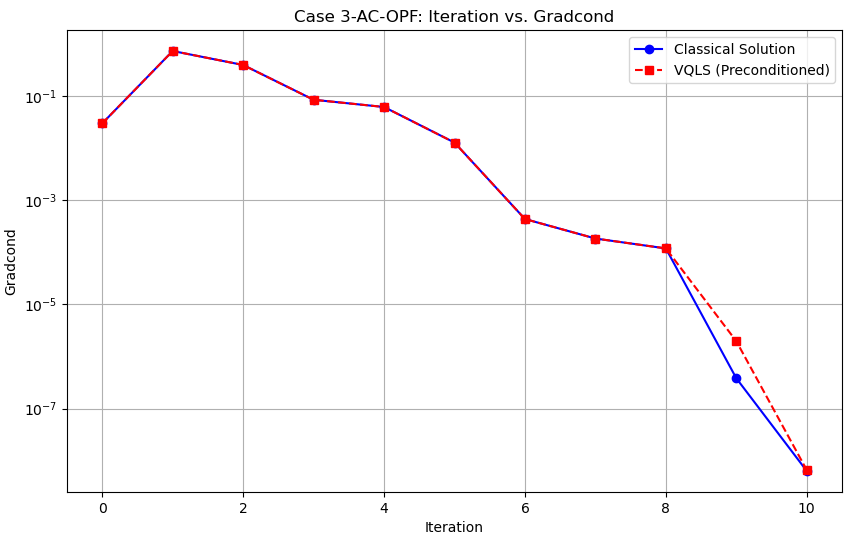}
    \caption{Case 3-AC-OPF:  \texttt{gradcond} v/s iterations}
    \label{fig:case3_ac_opf_iteration_vs_gradcond}
\end{figure}

Figure \ref{fig:case3_ac_opf_iteration_vs_gradcond} show the \texttt{gradcond} values for case 3 AC-OPF. The  \texttt{gradcond} values for the classical and VQLS methods are very close to each other in most iterations, suggesting that both methods perform the same. We can say that the VQLS method with left preconditioning technique performs comparably to the classical solution in this case.

Based on this limited simulation study, preconditioned  HHL and preconditioned VQLS do slightly worse in terms of convergence speed than the classical approach in MATPOWER for DC-OPF. For AC-OPF, the performance of preconditioned VQLS is close to that of the solver in MATPOWER.
% Limitations can be part of the discussions
\subsection{Limitations}

We could only simulate some smaller power cases in DC-OPF (case 3, 6ww, 9) and AC-OPF (case 3) using preconditioned VQLS and preconditioned HHL due to qubit requirement and numerical instability, other power cases \cite{matpower2011zimmerman} could not be solved effectively. 
%We plan to explore this more in future.

\section{Conclusion and Future Work}

In this paper, we propose hybrid quantum algorithms for solving OPF problems. Using a limited simulation study, we show that quantum approaches are comparable to classical approaches in terms of the accuracy of optimal costs and the number of iterations needed to reach
the convergence point. We also show that the growth in the condition number for DC-OPF and AC-OPF when computing the search direction can be addressed using a preconditioned HHL or VQLS. 
As part of future work, our research will focus on implementing quantum approaches for OPF on quantum hardware and exploring their scalability and performance in real-world power systems. We are also committed to tackling the remaining large-scale power cases by developing and implementing more efficient quantum approaches for OPF.

\section*{Acknowledgements:} This research is supported by an NSERC Quantum Alliance Grant, AB Innovates MIF Grant and was enabled in part by the Digital Research Alliance of Canada (https://alliancecan.ca/en).

\bibliography{references}
\bibliographystyle{unsrt} 

\end{document}